\documentclass[modern]{aastex63}
\usepackage[utf8]{inputenc}
\usepackage{amsmath}
\usepackage{framed}
\usepackage{url}       

\renewcommand{\twocolumngrid}{}

\newcommand\thecannon{\textsl{The Cannon}}

\newcommand\gaia{\textsl{Gaia}}
\newcommand\lamost{\textsl{LAMOST}}
\newcommand\apogee{\textsl{APOGEE}}
\newcommand\galah{\textsl{GALAH}}
\newcommand\method{TSP}
\newcommand\nstars{8,428}
\newcommand\reviewer[1]{#1}

\newcommand{\f}{\mathbf{f}}
\newcommand{\F}{\mathrm{F}}
\newcommand{\E}{\mathrm{E}}
\newcommand{\rr}{\mathbf{r}}
\newcommand{\m}{\widetilde}
\newcommand{\mv}{\mathbf{m}}
\newcommand{\fpred}{\m \f^\text{pred}}

\newcommand{\teff}{T_\mathrm{eff}}
\newcommand{\logg}{\log g}
\newcommand{\ppcs}{$\chi^2/\text{pixel}$}

\begin{document}\frenchspacing

\title{An unsupervised method for identifying $X$-enriched stars \\
directly from spectra: Li in \lamost{}}
\author[0000-0001-7339-5136]{Adam Wheeler}
\affiliation{Department of Astronomy, Columbia University, Pupin Physics Laboratories, New York, NY 10027, USA}
\author{David W. Hogg}
\affiliation{Center for Cosmology and Particle Physics, Department of Physics, New York University, 726 Broadway, New York, NY 10003, USA}
\affiliation{Center for Computational Astrophysics, Flatiron Institute, 162 5th Av., New York City, NY 10010, USA}
\affiliation{Max-Planck-Institut f\"ur Astronomie, K\"onigstuhl 17, D-69117 Heidelberg, Germany}
\author{Melissa Ness}
\affiliation{Department of Astronomy, Columbia University, Pupin Physics Laboratories, New York, NY 10027, USA}

\correspondingauthor{Adam Wheeler}
\email{a.wheeler@columbia.edu}

\begin{abstract}
Stars with peculiar element abundances are important markers of chemical enrichment mechanisms.
We present a simple method, tangent space projection (TSP),  for the detection of $X$-enriched stars, for arbitrary elements $X$, even from blended lines.
Our method does not require stellar labels, but instead directly estimates the counterfactual unrenriched spectrum from other unlabelled spectra.
As a case study, we apply this method to the $6708~$\AA{} Li doublet in \lamost{} DR5, identifying \nstars{} Li-enriched stars seamlessly across evolutionary state.
We comment on the explanation for Li-enrichement for different subpopulations, including planet accretion, nonstandard mixing, and youth.
\end{abstract}

\section{Introduction}
Recent years have seen many resolved-star spectroscopic surveys, and corresponding growth in the use of data-driven spectral models, e.g. \thecannon{} \citep{nessCannonDatadrivenApproach2015}, the DD-\emph{Payne} \citep{xiangAbundanceEstimates162019}, kernel principal component analysis approaches \citep{xiangEstimatingStellarAtmospheric2017}, and models using deep convolutional neural nets \citep{leungDeepLearningMultielement2019}.
These models have pushed astronomers into new regimes of precision \citep{jofreAccuracyPrecisionIndustrial2019}, and allowed us to infer evolutionary state, mass, and detailed abundances from low-resolution spectra \citep{hoMassesAges2302017, tingLargePristineSample2018, xiangAbundanceEstimates162019, wheelerAbundancesMilkyWay2020a, sandfordForecastingChemicalAbundance2020}.

Given a complete enough understanding of stellar atmospheres, the interstellar medium, the earth's atmosphere, and our instrumentation, data-driven spectral models would be unnecessary, but we are far from such total knowledge.
Most data-driven methods applied to stellar spectra (including all mentioned above) are concerned with supervised regression.
They use a set of spectra labelled with \emph{a-priori} atmospheric parameters, surface abundances, and reddening parameters, to calibrate a model which is then applied to unlabelled spectra.
These approaches are limited by the quantity and precision of training data, restricting the datasets, labels, and regimes to which they can be applied.
Furthermore, their reliance on labeled data means that they can be limited by the strong biases and  systematic errors introduced by physical models. 

In this work, we pursue an alternative approach.
Rather than estimate abundances, we endeavour only to identify stars that are highly enriched\footnote{Throughout this paper, we use ``enriched'' or $X$-enriched specifically to refer to stars with a high $X$ abundance \emph{relative to stars with similar spectra}---and hence parameters and abundances. A star which is $X$-rich may not be $X$-enriched, and vice versa.} in a particular element.
This relaxed goal permits approaches which require no labelled training data and minimal researcher input, but which retain high scientific payoff, since $X$-enriched stars often have undergone uncommon events.
Unusual abundance patterns may also signal birth in an accreted galaxy with an enrichment history different from the Milky Way's \citep[e.g.][]{hortaChemicalCompositionsAccreted2020, hawkinsChemicalNatureYoung2020, molaroLithiumBerylliumGaiaEnceladus2020,  vincenzoFallGiantChemical2019}.

We take advantage of the fact that stellar spectra, which are naturally expressed as high-dimensional flux vectors, are embeddable on (or near) a lower-dimensional manifold.
We censor the relevant absorption region of a spectrum, then use its neighbors on a local (euclidean) patch of the spectral manifold to impute the masked pixels.
This serves as a (possibly) couterfactual unenriched realization of the spectrum, against which we examine the residuals to identify ``unexpectedly'' strong absorption.

This work is related to ideas in manifold learning and nonlinear dimensionality reduction, especially  local linear embedding \citep{roweisNonlinearDimensionalityReduction2000} and isomap \citep{tenenbaumGlobalGeometricFramework2000}.
Hessian local linear embedding \citep{donohoHessianEigenmapsLocally2003} shares with our method the use of singular value decomposition to estimate the tangent space of the data-manifold.
Unlike these methods, we never explicitly construct global nonlinear coordinates since all of our calculations can be performed in a small patch on the spectral manifold.
While we leverage the embeddability of spectra in a low-dimensional space, we never construct a continuous low-dimensional representation.

This is not the first unsupervised model deployed on a large spectral survey.
\citet{feeneySSSpaNGStellarSpectra2019} used a fully probabilistic nonparametric model to characterize the spectra of \apogee{} red clump stars, denoising them by a factor of a few and demonstrating the mutual information present in features of elements belonging to the same nucleosynthetic family.
Though their model could, in principle, be used to impute masked pixels, it is too expensive to deploy across all wavelengths of all observed stars.
\citet{zerjalGALAHSurveyLithiumstrong2019} used a simple nearest-neighbor method to identify Li-enriched KM dwarfs in \galah{}.
\citet{cotarGALAHSurveyCharacterization2020} used autoencoders (a family of neural-net architectures) to identify emission stars.

We turn our attention to stars that are enriched in Li.
Li-7 burns at a mere $2.5 \times 10^6~\mathrm{K}$ \citep{bodenheimerStudiesStellarEvolution1965}, and is depleted at all stages of stellar evolution, but thought to be replenished in myriad ways. 
Its low ionization potential means that in stellar atmospheres Li exists mostly as Li-II, which is not detectable.
Li-I's strongest feature (the 6708 \AA{} doublet) is thus very weak and $\teff$-sensitive.
It is also sensitive to non LTE (local thermodynamic equilibrium) effects \citep{lindDeparturesLTENeutral2009}.
All this means that Li can be challenging to study with physical models.


\section{Data} \label{sec:data}
We use \reviewer{the LAMOST DR 5 \citep{dengLAMOSTExperimentGalactic2012a, zhaoLAMOSTSpectralSurvey2012} AFGK sample}, which contains $5.3 \times 10^6$ low-resolution ($R=1800$) spectra of $4.3 \times 10^5$ unique stars.
It covers wavelengths from 3800 \AA{} to 9000 \AA{}.
We analyze repeat observations independently with \method{}, but stack the resulting residuals before identifying strong absorption.
We pre-treat all LAMOST spectra by interpolating to a common rest-frame wavelength grid, then applying the approximate continuum normalization first used in \citet{hoLabelTransferAPOGEE2017}, that is, dividing the spectrum by a itself smoothed with a 25 \AA{} Gaussian kernel.
While this transformation distorts broad features, it is applied homogeneously across all spectra and thus will not introduce biases.
We impute any pixels with normalized flux, $f$, outside of $0 \leq f \leq 2$ or with uncertainty greater than $\frac{1}{2}$ by setting $f=1$, and setting the associated uncertainty to \texttt{inf}.

\section{Methods}
The inputs to our algorithm are the following:
\begin{itemize}
\item The data, assumed to be a set of uniform vectors.  In this work these are rest-frame spectra, interpolated onto a common wavelength grid.
\item The reference set, a subset of the data well-distributed throughout the underlying parameter space \footnote{In machine learning terminology, this might be called the training set. In the L-ISOMAP dimensionality reduction algorithm \citep{desilvaGlobalLocalMethods2002}, these data are called ``landmarks''.}
\item Integers $k$ and $q$ which specify the number of neighboring data points to use and the dimensionality of the manifold, respectively.
\item The components of each data point that are of interest.  For stellar spectra, these are the pixels containing the spectral feature(s) under investigation. We will refer to data as \emph{censored} when these components are dropped. We refer to the components themselves as \emph{masked}.  We take $n$ to be the number of unmasked pixels, and $m$ to be the number of masked pixels.
\end{itemize}
We take each spectrum to have flux uncertainty described exactly by a multivariate normal distribution with known covariance.
The most naive version of this algorithm requires complete data.  
As discussed in section \ref{sec:data}, we obtain complete data by imputing bad pixels with $f=1$.
Note, however, that \method{} could in principle itself be used to impute bad pixels more robustly. 

Ideally, the reference set would include every spectrum available (perhaps excluding those with strong absorption features, see discussion in section \ref{sec:discussion}).
Using a random subset of spectra instead speeds up computation.

We need to know the expected profile of the spectral feature under investigation to identify enriched stars after imputing.
For all but the strongest absorption lines in low-resolution spectra, knowing the instrument resolution is sufficient, since the line's observed profile is dominated by the line spread function.


Figure \ref{fig:algo} presents an overview of the algorithm, section \ref{sec:imputing} goes into detail, and \ref{sec:flagging} describes our approach to identifying enriched stars from imputed residuals.

\begin{figure}
    \begin{framed}
    For each spectrum, $\f$:
        \begin{enumerate}
            \item Compare $\f$ to all spectra in the reference set (with the region of the absorption feature masked) to find its $k$ nearest neighbors.
            \item Compute a basis for the $q$-hyperplane that captures as much variance amongst the neighbors as possible, the approximate tangent space.
            \item Impute the masked pixels by projecting $\f$ onto the approximate tangent space.
            \item Determine if the residuals corresponds to excess absorption.
        \end{enumerate}
    \end{framed}
    \caption{Algorithm summary.}
    \label{fig:algo}
\end{figure}

\subsection{Tangent Space Projection} \label{sec:imputing}
Stated briefly, \method{} is imputation of masked data via local principal component regression. 
Here we detail how to use \method{} to predict masked spectral pixels of an arbitrary target spectrum, using a reference set of randomly-selected unlabeled spectra.
The reference set and target spectra are assumed to be from the same instrument and interpolated to the same wavelength grid.
Take the target spectrum's $\lambda^\mathrm{th}$ unmasked pixel to be $f_\lambda$.
First find the $k$ nearest neighbors in the reference set (leaving out the masked portion of the spectra), i.e. those that minimize the euclidean distance (L2 norm), 
\begin{equation}
d = \sqrt{\sum_{\lambda=1}^{n}(f_\lambda - f_\lambda^\mathrm{ref})^2} \quad,
\end{equation} 
for a reference spectrum whose $\lambda^\mathrm{th}$ pixel is $f_\lambda^\mathrm{ref}$.
We exclude those reference spectra with missing data in the masked region.

Take $\F$ to be the $k \times (n + m)$ matrix whose rows are the uncensored neighboring spectra in the zero-mean basis (with the neighborhood mean subtracted from each row).
Next, calculate the $q$-hyperplane that captures as much variance as possible amongst the neighbors (using their full, uncensored spectra), that is the first $q$ principal components of $\F$.
Let $\E$ be the $q \times n$ matrix whose rows are the first $q$ censored $n$-pixel eigenspectra, and let $\m\E$ be the $q \times m$ matrix whose rows are the $m$ masked pixels of the first $q$ eigenspectra.
\reviewer{Together, $\E$ and $\m\E$ are an estimate of the tangent space of the spectral manifold. In the context of Equation \ref{eq:proj}, below, each is a design matrix whose rows can be thought of as features.}

Let $\f$ be the $n$-pixel (censored) target spectrum in the zero-mean basis, and let $\Sigma$ be the $n \times n$ covariance matrix describing the uncertainty in $\f$.
We can predict the \reviewer{masked pixels}, $\fpred$, by projecting $\f$ onto the row space of $\E$:
\begin{equation}\label{eq:proj}
\fpred = \m\E \underbrace{(\E^T \Sigma^{-1} \E)^{-1} \E^T \Sigma^{-1} \f}_\text{linear least-squares coeffs} \quad.
\end{equation}
\reviewer{The marked part of this equation is uncertainty-weighted linear least-squares regression (\citealp{hastieElementsStatisticalLearning2009} chapter 3 and \citealp{hoggDataAnalysisRecipes2010} provide comprehensive reviews); it is a representation of $\f$ in $q$-dimensional local coordinates. That representation is then mapped to the masked pixels through multiplication by $\m\E$.}
$\fpred$ is the maximum-likelihood estimate for the $m$ masked pixel values given $\f$ and $\Sigma$, assuming that the ground truth spectrum lies on the hyperplane spanned by the first $q$ eigenspectra.
Treating the eigenspectra, $\m\E$ and $\E$, as fixed, the covariance describing the uncertainty on $\m \f^\text{pred}$ is 
$\m\Sigma_\text{pred} = \m\E (\E^T\Sigma^{-1} \E)^{-1} \m\E^T$.
In practice, we neglect this uncertainty since when $n \gg m$ (i.e. when a small fraction of all pixels are masked) it is subdominant to the measurement uncertainty in the masked pixels, $\m \f$.

\subsection{Identifying Strong Absorption in the Residuals} \label{sec:flagging}
\reviewer{Having used the unmasked pixels to predict the masked ones}, we can examine the residuals between the measured and predicted masked pixels to identify unusually strong absorption features and eliminate contaminants.
We use a simple likelihood ratio test, i.e. matched filtering.

\reviewer{Let $\m \rr$ be the residuals of the prediction of the fluxes of the masked pixels, $\m \rr = \m \f - \m \f ^\text{pred}$, and let $\mathrm{C}$ be the covariance matrix describing its uncertainty.}
Consider a family of models of the form $\alpha \mv$, where $\alpha$ is a scalar amplitude and $\mv$ is a fixed profile.  
Since the uncertainty on $\m\f^\text{pred}$ is small, $\mathrm{C}$ is the diagonal matrix whose entries are the squares of the per-pixel measurement uncertainty of $\m\f$.
The value of $\alpha$ which maximizes the likelihood is 
\begin{equation}
    \alpha^* = \frac{\mv^T\mathrm{C}^{-1}\m\rr}{\mv^T \mathrm{C}^{-1} \mv} \quad,
\end{equation}
with uncertainty
\begin{equation}
   \sigma_{\alpha^*} = (\mv^T\mathrm{C}^{-1}\mv)^{-1/2} \quad.
\end{equation}

For a given $\mv$, the optimal-amplitude likelihood is
\begin{equation} 
    \log p(\m\rr | \alpha^* \mv)  = \frac{1}{2}\frac{(\mv^T\mathrm{C}^{-1}\m\rr)^2}{\mv^T \mathrm{C}^{-1} \mv} + \text{const} \quad.
\end{equation}
If $\mv$ is a line model with equivalent width $w$, $\alpha^*w$ can be interpreted as best-fit excess equivalent width (EEW), equivalent width in excess of that found in a counterfactually unenriched spectrum.

\begin{figure}
    \centering
    \includegraphics{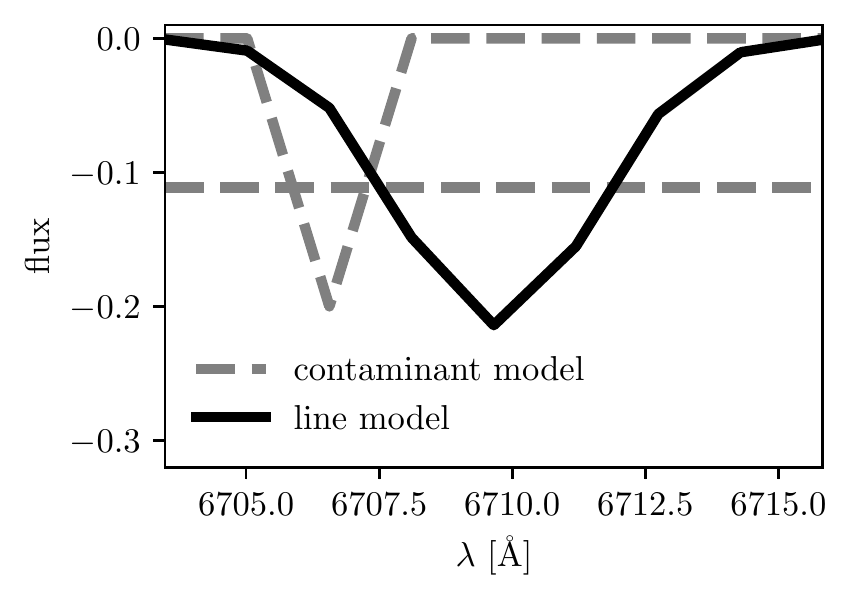}
    \caption{Our line model (black) and contaminant models (grey, dashed). Only one single-pixel contaminant model is shown for clarity.}
    \label{fig:contaminantmodels}
\end{figure}

We calculate the likelihood, amplitude (EEW), and amplitude uncertainty for a Gaussian (in wavelength) absorption feature with width given by the instrument resolution, as well as for two contaminant models: uniform residuals, as might arise from a poorly modelled continuum, and residuals with a single non-zero pixel.
Our line model and contaminant models are plotted in Figure \ref{fig:contaminantmodels}.
We require the contaminant models to be less likely than the line model to identify a star at Li-enriched (see section \ref{sec:results}).

\subsection{Hyperparameter Selection}\label{sec:hyperparams}
\begin{figure}
    \centering
    \includegraphics[width=0.6\textwidth]{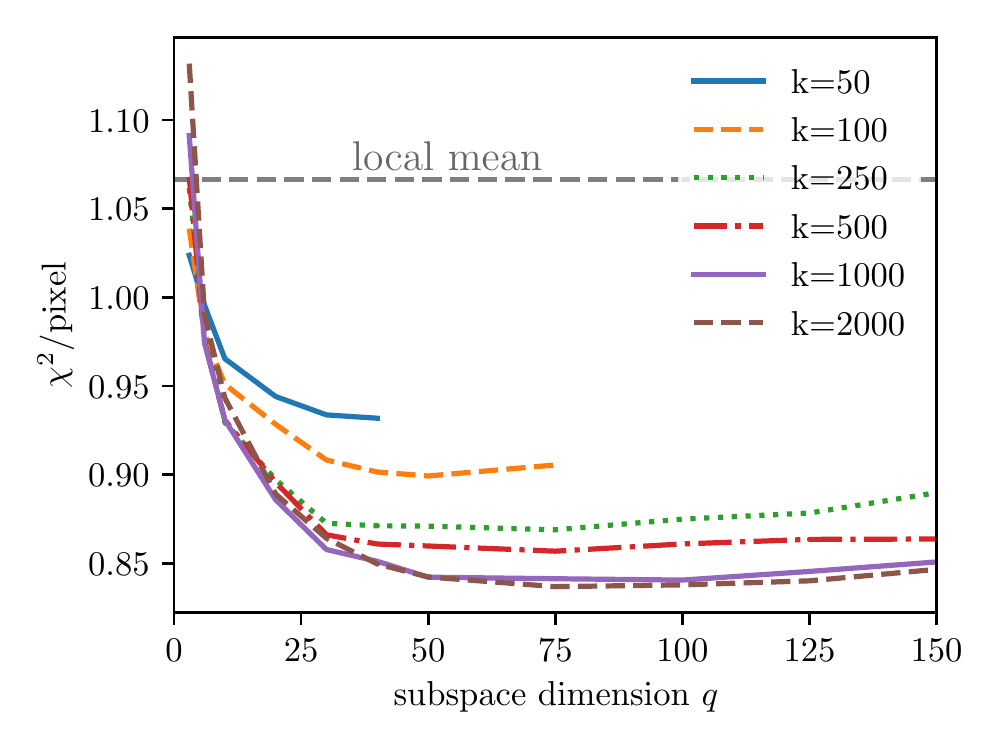}
    \caption{Mean squared error of the predicted spectral region as a function of subspace dimension, $q$, for different neighborhood sizes, $k$. The dashed grey line marks the best $\chi^2/\text{pixel}$ ($k = 3$) achieved by using the local mean spectrum as a prediction. We use $k=1000$, $q=50$ for our analysis.}
    \label{fig:MSEq}
\end{figure}

To inform the choice of hyperparameters $k$ and $q$, we evaluate \method{}'s predictive accuracy for arbitrary spectral regions.
We take 1000 random \lamost\ spectra and predict a different randomly-selected contiguous region of 100 pixels (roughly an order of magnitude larger than a mask for a single line) for each.
Figure \ref{fig:MSEq} shows the per-pixel $\chi^2$  of the masked pixels as a function of subspace dimension, $q$, for different neighborhood sizes, $k$.
Large values of each ($k \gtrsim 250, q \gtrsim 25$) give better predictions of the masked pixels, which is consistent with the large number latent parameters found to meaningfully describe \lamost{} spectra by \citep{xiangEstimatingStellarAtmospheric2017}.
The nominal measurment uncertainty is saturated for nearly all values of $k$ and $q$, indicating that it may be overestimated\reviewer{, although our continuum normalization procedure introduces correlations between nearby pixels that may contribute to this.}
Using the mean of the $k$ nearest neighbor spectra as a prediction, while simpler than \method{}, is not as effective.
The horizontal dashed line in Figure \ref{fig:MSEq} shows the predictive accuracy for $k=3$, the neighborhood size for which a local average performs best.
\method{} achieves better predictive accuracy for nearly all hyperparameter values considered.

\begin{figure}
    \centering
    \includegraphics[width=0.6\textwidth]{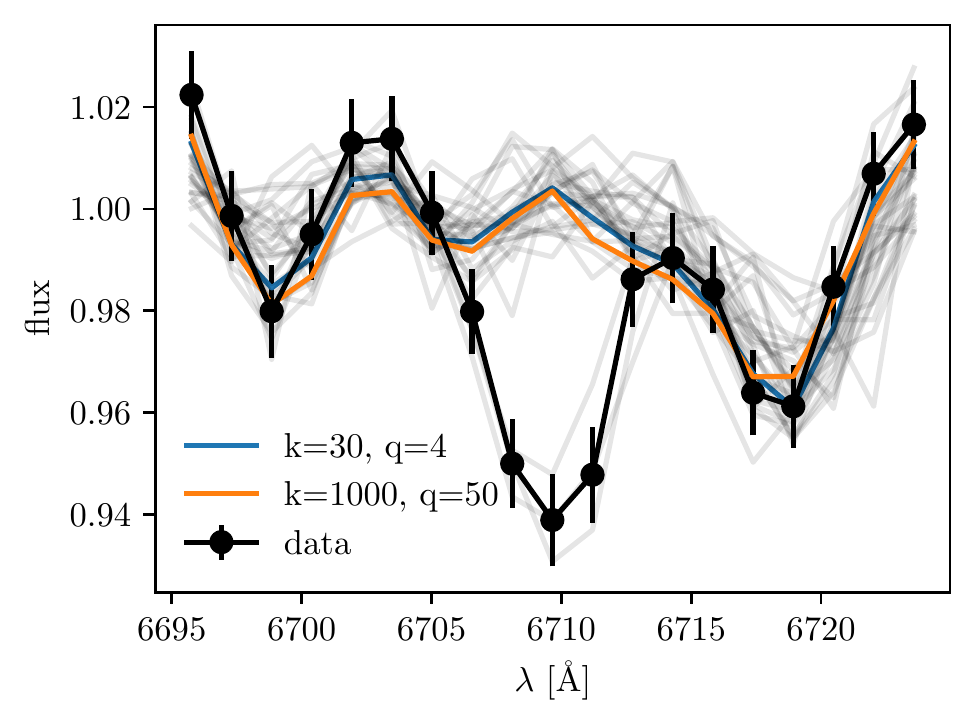}
    \caption{The prediction of the Li absorption region in a spectrum of \lamost{} J062739.73+463634.4, 
    a Li-enriched giant. The imputed spectrum for $k=30, q=4$ is very similar to that for $k=1000, q=50$.  Neither predict the absorption seen in the data. The 30 nearest spectral neighbors are shown in gray.}
    \label{fig:hiqloq}
\end{figure}

For our case-study, we choose $k=1000$ and $q=50$.
We remark, however, that predictive accuracy in the Li absorption region is not especially sensitive to this choice, as demonstrated by Figure \ref{fig:hiqloq}, which shows the masked portion of a spectrum exhibiting strong Li absorption with \method{}'s predictions for $q=4$ and $q=50$.
\method{} imputes these pixels very similarly regardless of our choice of $q$.
Note that TSP predicts no absorption, even though some neighbors have absorption similar to that of the spectrum being imputed.
\reviewer{This is because both Li absorption features have been masked, and the unmasked pixels do not contain correlated spectral features.}

We found that increasing the size of the reference set decreased the \ppcs{}, but changed the preferred values of $k$, as expected.
We use a reference set of 30,000 randomly chosen spectra, but remark that a larger reference set would presumably marginally decrease predictive \ppcs{} at a greater computational cost.
Using a reference set composed exclusively of spectra with a high signal to noise ratio ($S/N > 100$) results in poor predictive accuracy.
\reviewer{Examination of stellar parameters for repeat observations of the same star did not reveal any S/N-dependent bias in stellar parameters
However, both continuum-normalized and un-normalized spectra of repeat observations have small S/N-dependent systematics, indicated that both the \lamost{} reduction pipeline and our continuum-normalization introduce biases which may prevent a reference set of all high-S/N spectra from performing well on low-S/N spectra. 
}

\section{Results} \label{sec:results}
\begin{deluxetable*}{llp{10cm}} \label{tab:cat}
\tablecaption{ Catalog schema.  Data available at \dataset[10.7916/d8-3ap9-qe35]{https://doi.org/10.7916/d8-3ap9-qe35}.
}
\label{tab:schema}
 \tablehead{
 \colhead{column name} & \colhead{type} & \colhead{description}
 }
 \startdata 
\texttt{designation} & string & \lamost\ unique star identifier  \\
\texttt{source\_id} & string & \gaia{} identifier  \\
\texttt{diff} & float array & flux residuals for each masked pixel  \\
\texttt{ivar} & float array & inverse variance in the flux residuals\\
\texttt{max\_best\_fit\_chi2} & Float & largest whole-spectrum $\chi^2$ for any observation of this star \\
\texttt{isline} & bool array & true when the line model is more likely than any contaminant model  \\
\texttt{likelihoods} & float array & the likelihood values (up to an additive constant) of the line model and each contaminant model \\
\texttt{EEW} & float & excess equivalent width [\AA]\\
\texttt{EEW\_err} & float & uncertainty in the EEW, $\sigma_\text{EEW}$ [\AA]\\
\texttt{enriched} & bool & true if the star is Li-enriched per definition in \ref{sec:results}\\
\texttt{ra} & float & right ascention  \\
\texttt{dec} & float & \lamost\ unique star identifier  \\
\texttt{teff} & float &  $\teff$ [K] from \lamost{} (mean of observations)\\
\texttt{logg} & float &  $\logg$ from \lamost{} (mean of observations)\\
\texttt{feh} & float &   [Fe/H] from \lamost{}  (mean of observations)\\
\texttt{snrr} & float & \lamost{} $r$-band $(S/N)$ (largest of all observations)\\
\enddata
\end{deluxetable*}
 
\begin{figure}
    \centering
    \includegraphics[width=\textwidth]{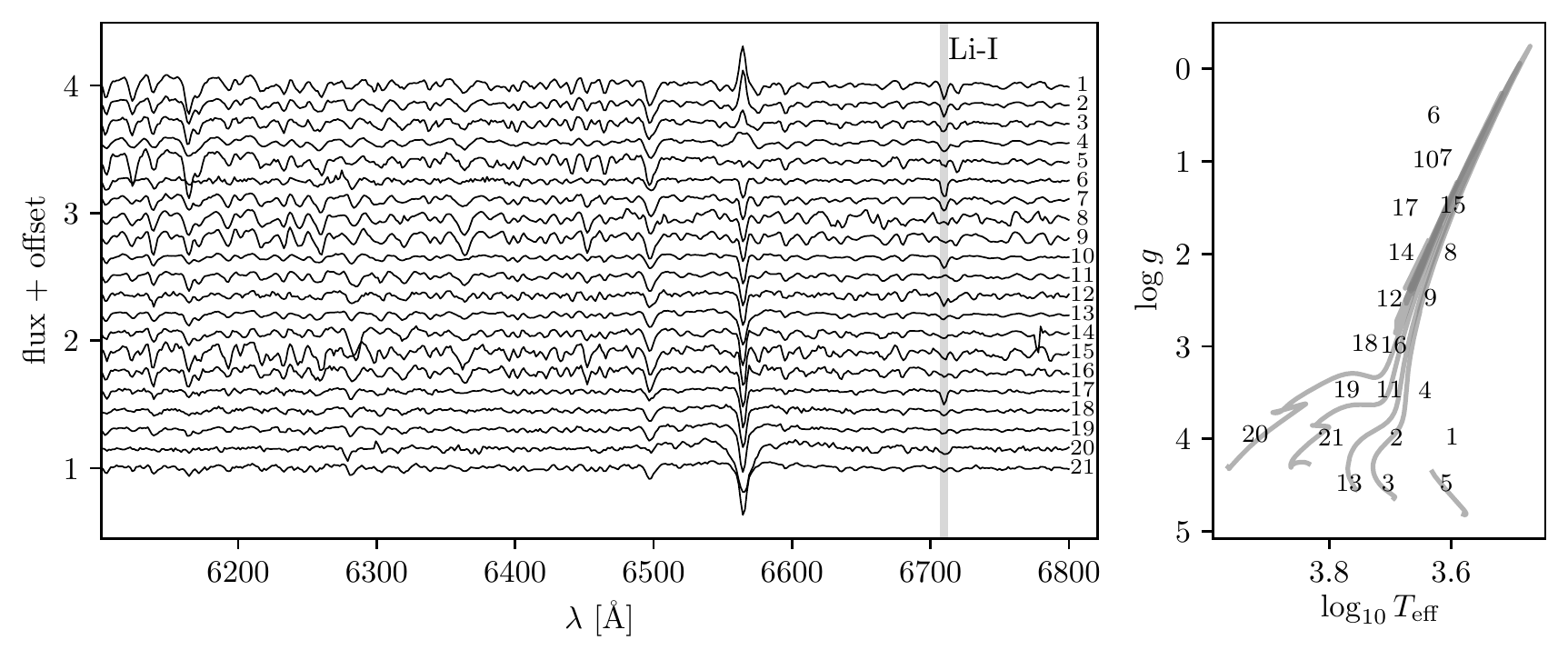}
    \caption{\textbf{left:} Portions of 21 spectra of Li-enriched stars spaced roughly evenly on the Kiel diagram and sorted by $H_\alpha$ (6565 \AA) amplitude. \textbf{right:} their positions on the Kiel diagram, with mass tracks for 2 $M_\odot$, 1.5 $M_\odot$, 1 $M_\odot$, 0.8 $M_\odot$, and 0.5 $M_\odot$ solar metallicity stars (details in text).}
    \label{fig:spectra}
\end{figure}

We apply our method to the 6708 \AA{} feature, masking wavelengths in the range $6703 ~\text{\AA{}}$ --  $6717~\text{\AA{}}$ (vacuum). 
Table \ref{tab:cat} shows the output quantities for each star.
We mask, but do not use, the 6106 \AA{} Li feature.
To consider stars to be Li-enriched, we require the detection of an absorption feature with an EEW of at least 0.15 \AA{} with an uncertainty excluding null EEW at the $3\sigma$ level ($\sigma_\text{EEW} < 0.05$ \AA), and that the absorption line model fit the data better than any contaminant model.
\reviewer{Of these, 147 stars have Li abundances available in \emph{GALAH} DR3 \citep{buderGALAHSurveyThird2020}.  
They have a mean [Li/Fe] of 1.7, with a standard deviation of 0.5, and 74\% have $\text{[Li/Fe]} > 1.5$, the traditional threshold.}
When evaluating the population properties of Li-enriched stars, we throw out those stars whose uncertainties make such a detection impossible (those for with EEW uncertainty greater than 0.05 \AA{}, about 40\% of all \lamost{} stars).
Hereafter, we refer to stars with $\sigma_\text{EEW} < 0.05$ as \emph{candidates} (and those that are both candidates and have $\text{EEW} > 0.15$ simply as \emph{Li-enriched}).
Within the set of candidates, there is no dependence of EEW on $S/N$.
Figure \ref{fig:spectra} shows portions of 21 Li-enriched spectra on different parts of the Kiel diagram.

\begin{figure}
    \centering
    \includegraphics{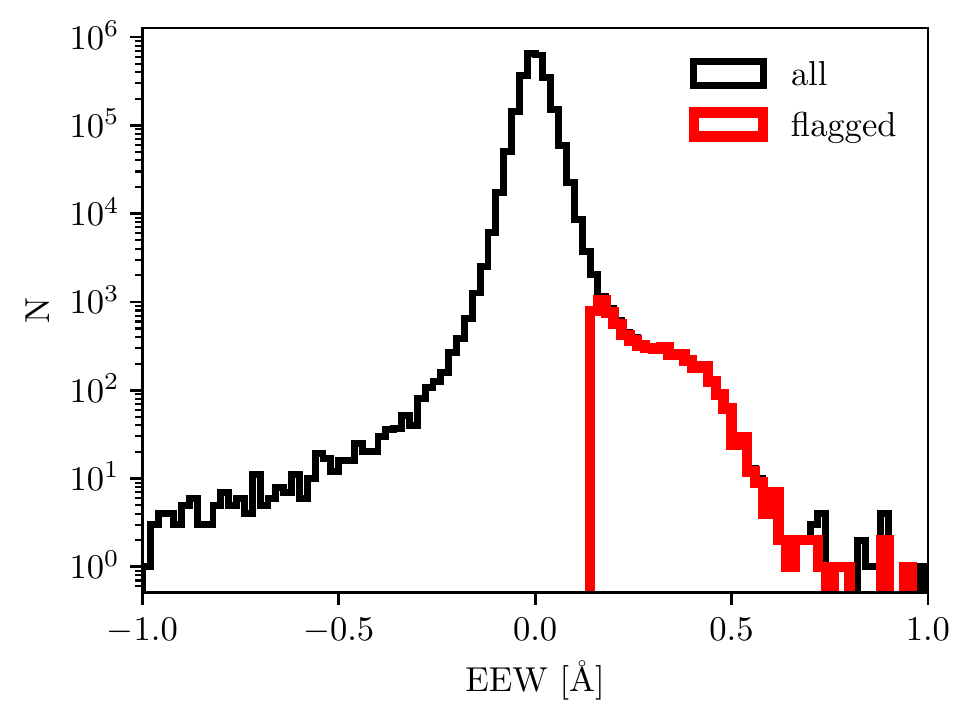}
    \caption{The distribution of excess equivalent widths for all candidates (those with $\sigma_\text{EEW} < 0.05 \text{\AA{}}$) and for Li-enriched stars (candidates with $\text{EEW} > 0.15 \text{\AA{}}$).}
    \label{fig:EEWhist}
\end{figure}

Figure \ref{fig:EEWhist} shows a histogram of the EEW values for all candidate stars, and all Li-enriched stars.
Nearly all stars with $\text{EEW} > 0.15 ~\text{\AA{}}$ have their masked-pixel residuals best matched by the line model (see section \ref{sec:flagging}), except at $\text{EEW} \gtrsim 0.7~\mathrm{\AA}$, where a significant fraction of spectra have continuum offsets.
\reviewer{The width of the EEW distribution is narrower than naively expected given the nominal \lamost{} measurement uncertainty, again indicating that it is overestimated.}
There is an uptick of in the number of stars with EEW around 0.2 -- 0.6 \AA{}, formed by the excess of Li-enriched stars.
\reviewer{There are \nstars{} Li-enriched stars, and 2,147 candidate stars best fit with the Li line model with $\text{EEW} < -0.15~\text{\AA}$, suggesting a contamination rate of up to 25\% (the presence of any Li-depleted stars will inflate this figure).}

\begin{figure}
    \centering
    \includegraphics[width=0.95\textwidth]{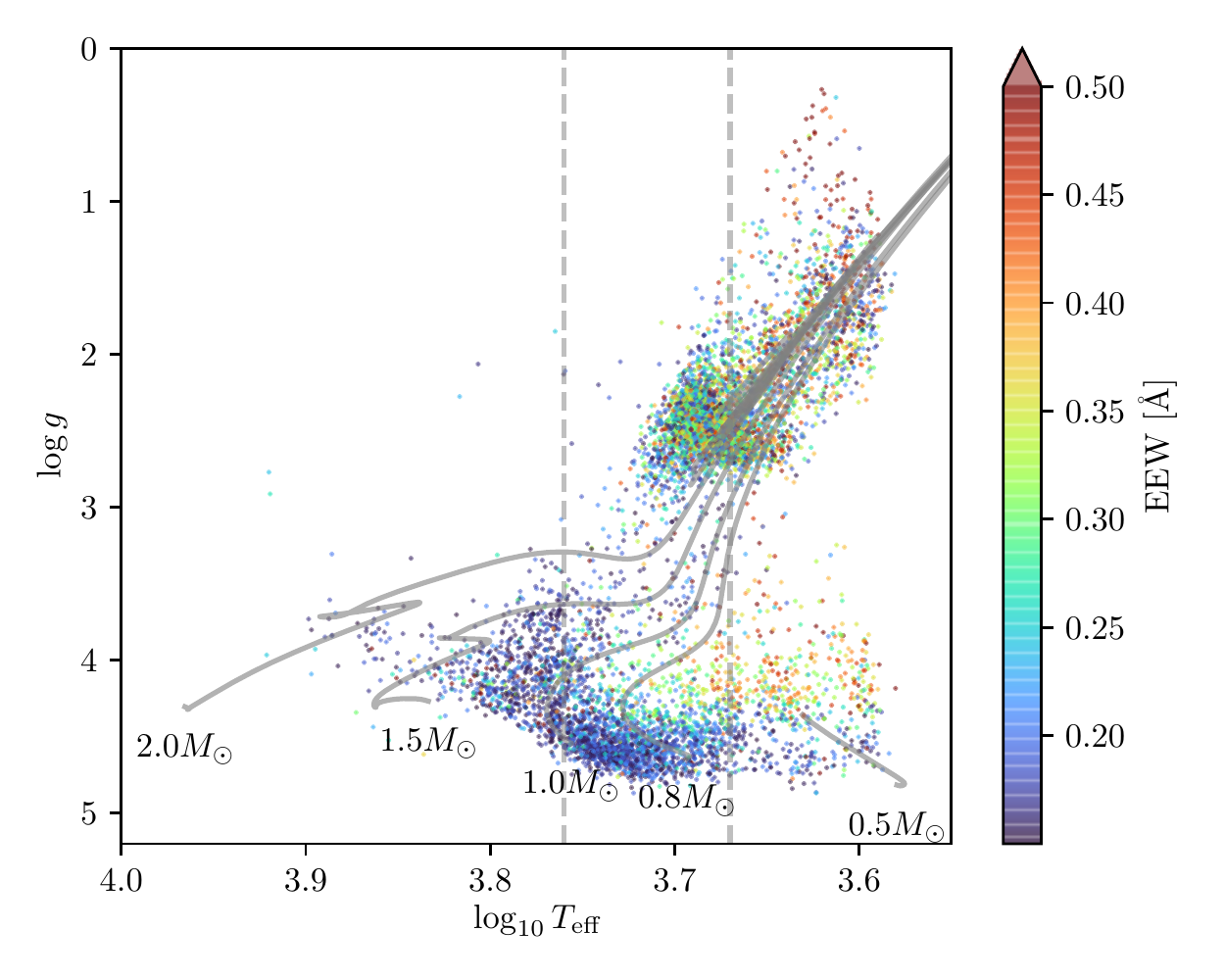}
    \includegraphics[width=0.44\textwidth]{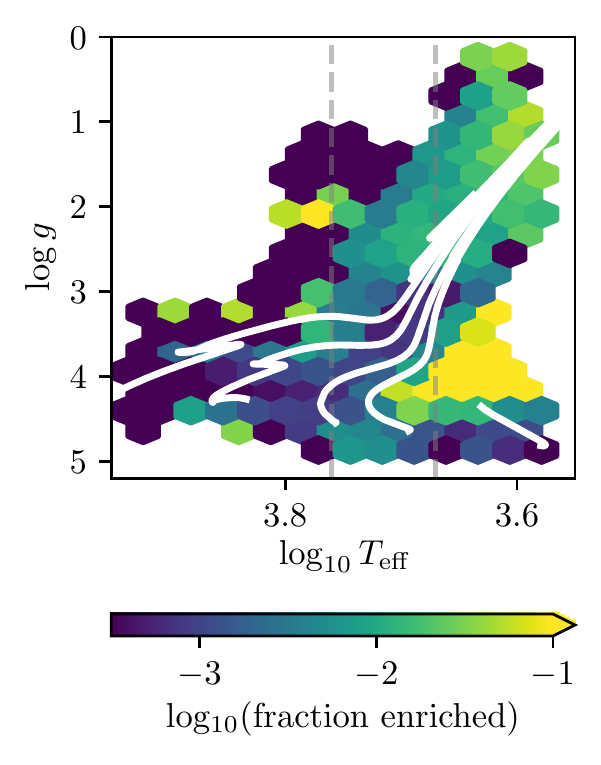}
    \includegraphics[width=0.44\textwidth]{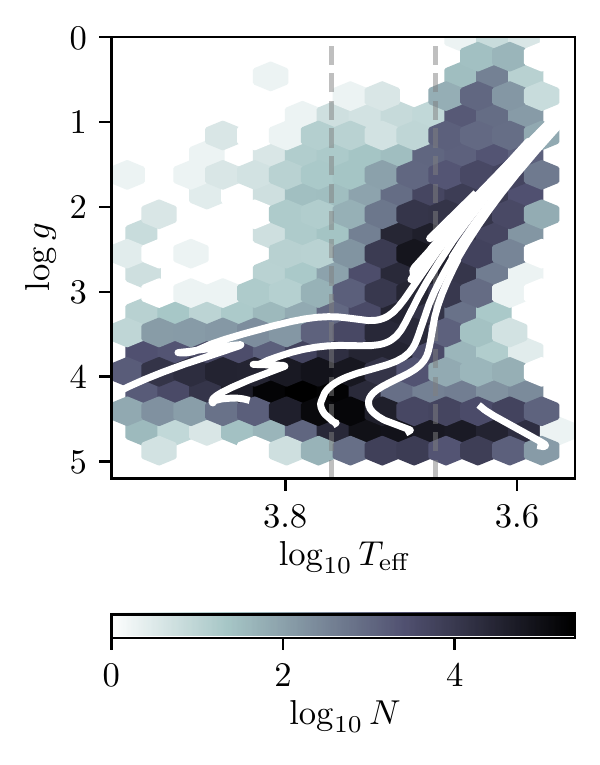}
    \caption{\textbf{top:} The \nstars{} Li-enriched stars plotted on the Kiel diagram with MIST solar-metallicity mass tracks. 
    \textbf{bottom left:} fraction all stars with $\sigma_\text{EEW} < 0.05$ \AA{} with are Li-enriched.
    \textbf{bottom right:} the number of of stars with $\sigma_\text{EEW} < 0.05$ \AA{}.
    Dashed vertical lines are the same as in Figure \ref{fig:pleiades}.  \reviewer{They mark the range over which young stars are Li-enriched.}}
    \label{fig:kiel}
\end{figure}

There are no obvious spatial trends in the fraction of Li-enriched stars; there are only trends in stellar parameters and evolutionary state.
Figure \ref{fig:kiel} shows the \nstars{}  Li-enriched stars in the $\logg$--$\teff$ plane (the Kiel diagram), along with the occurrence fraction of Li enrichment, and the number of candidates.
Plotted for comparison are solar-metallicity mass tracks from MESA isochrones and stellar tracks\footnote{MIST version 1.2.  Mass tracks generated with initial $v/v_\text{crit}$ set to $0.4$.} (MIST; \citealp{dotterMESAIsochronesStellar2016a, choiMesaIsochronesStellar2016, paxtonModulesExperimentsStellar2011, paxtonModulesExperimentsStellar2013, paxtonModulesExperimentsStellar2015}), 
Li enrichment is especially prevalent among pre-main sequence stars, stars near the zero-age main sequence (ZAMS), especially at larger $\teff$, subgiant branch stars, and red giants at and above the red clump.  We go into detail in the sections below.

\subsection{Planet Accretion on the Subgiant Branch}\label{sec:planetaccretion}
Presumably, stars occasionally consume their satellites.  
These events may enrich the star's photosphere, especially when the satellites are large.
Accretion of sub-stellar companions has been identified many times as a Li-enrichment mechanism \citep{alexanderPossibleSourceLithium1967, siessAccretionBrownDwarfs1999, villaverOrbitalEvolutionGas2009,adamowBD487402012, carlbergObservableSignaturesPlanet2012, ohKronosKriosEvidence2018}, and detection of cometary material accreted by white dwarfs has a long history \citep[e.g.]{vanmaanenTwoFaintStars1917,weidemannAtmosphereWhiteDwarf1960, zuckermanMetalLinesWhite2003, keplerNewWhiteDwarf2016}.
Recently, \citet{soares-furtadoLithiumEnrichmentSignatures2020} found that, while photospheric Li will become unobservable within roughly $10$ years for red giants, which have deep convective envelopes, it has a lifetime of up to 1 Gyr  for 1.4 -- 1.6 $M_\odot$ stars on the subgiant branch and near the main sequence turn-off.
Compounded with the fact that stars on the subgiant branch are expanding rapidly, this suggests that planet accretion could account for a significant fraction of Li-enriched subgiants.

Figure \ref{fig:kiel} shows a moderate enrichement fraction ($\sim 0.1 \%$) stars near the 1.5 $M_\odot$ subgiant branch.
Neither the data-driven abundances of \citet{xiangAbundanceEstimates162019} (Na, Mg, Al, Si, Ca, Ti, Cr, Mn, Co, Ni, Cu, Ba) nor \citet{wheelerAbundancesMilkyWay2020a} (O, Sc, Eu, Mg, Al, Mn, Ba) show a significant difference in mean abundances between Li-enriched and normal subgiants.
Stars that have recently accreted planets are speculated to be especially enhanced in refractory elements, those with a high condensation temperature \reviewer{(e.g. \citealp{ramirezAccurateAbundancePatterns2009,melendezPeculiarSolarComposition2009, melendezSolarTwinPlanet2017})}.
We find no evidence for a positive trend between condensation temperature and comparative enrichment of Li-enriched subgiants over Li-normal subgiants.
In fact, a weak negative trend is present---Li-enriched stars are preferentially depleted in refractory material, with the exception of Ba.

\subsection{Li-enriched Red Giants}

More than half the stars we identify as Li-enriched are giants.
Of the candidate stars, 4459 have $\logg < 2.7$, consistent with the roughly 1\% of red giants thought to be Li-enriched (traditionally defined as $A(\mathrm{Li}) > 1.5$; e.g. \citet{gaoLithiumrichGiantsLAMOST2019}, \citealp{caseyTidalInteractionsBinary2019}).
As seen in Figure \ref{fig:kiel}, they span temperatures $4000~\text{K} \lesssim \teff \lesssim 5600~\text{K}$ ($3.6 \lesssim \log_{10} \teff~[\text{K}] \lesssim 3.75$).

The first Li-enriched red giant is often considered to be that identified by \cite{wallersteinGiantUnusuallyHigh1982}, although a Li-enriched asymptotic giant branch (AGB) star was reported four decades earlier by \citet{mckellarIntense6708Resonance1940}.
As mentioned in section \ref{sec:planetaccretion}, sub-stellar companion accretion is unlikely to explain a significant fraction of Li enrichment for red giants.
Cosmic ray spallation can account for some Li-enrichment \citep{burbidgeSynthesisElementsStars1957, mitlerOriginLightElements1964}, but only a small fraction, and does not produce the observed isotope ratio \citep{reevesOriginLightElements1994}.
Mass-transfer from an AGB star companion can account for Li-enrichment in those that have one, but not in isolated giants.
Li-enriched giants are not preferentially found in binaries \citep{adamowTrackingAdvancedPlanetary2018}.
Li-enriched material is produced by classical novae \citep{starrfieldDustFormationNucleosynthesis1997, molaroHighlyEnriched7Be2016} and Type II supernovae \citep{dearbornShockingDevelopmentLithium1989}, but it is unclear whether, and in what quantity, this material can be accreted onto giant stars.
 
\citet{cameronOriginAnomalousAbundances1955} and \citet{cameronLithiumSPROCESSRedGiant1971} first suggested that surface Li could be enhanced if Be-7 is transported via convection from a depth at which PP-II fusion is occurring and decays via electron capture to Li-7.
For the Be to be transported to a cool layer of the star before it decays into Li, the convection timescale must be faster than its decay timescale, thought to be 50 -- 100 yr \citep{cameronOriginAnomalousAbundances1955}.
While this process is thought to occur as originally suggested in AGB stars \citep{deepakStudyLithiumrichGiants2019, singhSurveyLirichGiants2019a}, driven solely by convection \citep{smithSynthesisLithiumSprocess1989} or thermal pulsations (e.g. \citealp{habingAsymptoticGiantBranch2011} chapter 1), it must be augmented with an additional mixing mechanism for RGB stars.
Two possibilities are internal rotation \citep{sweigartMeridionalCirculationCNO1979, fekelLithiumRapidRotation1993, charbonnelConsistentExplanation12C1995}, and thermohaline mixing \citep{sackmannCreation7LiDestruction1999, charbonnelNatureLithiumRich2000, denissenkovThermalStabilityRotating2003, lattanzioNumericalTreatmentDependence2015}.
\citet{caseyTidalInteractionsBinary2019} found that tidal interactions were the likely culprit, while \citet{martellGALAHSurveyLithiumrich2020} found that at least two mechanisms are likely in effect.
Recently, \citet{kumarDiscoveryUbiquitousLithium2020} found that a ubiquitous process is in operation for all low-mass stars, on the basis of the high Li abundances in the red clump. 

\reviewer{
Extremely few of our Li-enriched giants are Ba-enriched (see section \ref{sec:ba}), which indicates that thermally-pulsating AGB are not contributing significantly to the Li-enriched sample.
Early AGB stars (E-AGBs), which may not experience significant dredge-up are more difficult to observationally distinguish from red giants. 
That we see a much higher Li-enrichment rate at and above the red clump than below is suggestive that E-AGBs may make up a significant fraction of the Li-enriched sample.
On the other hand, our enrichment rate is similar to past studies, suggesting that we are identifying red giants.
}

We see no evidence on the basis of the $\logg$ and $\teff$ distributions of enriched stars that Li-enrichment is more common in the red clump (RC) than on the RGB, in sharp contrast with \citet{deepakStudyLithiumrichGiants2019, martellGALAHSurveyLithiumrich2020, caseyTidalInteractionsBinary2019}.
This may be partially attributable to our direct use of absorption feature, rather than calculating the abundance from the equivalent width.
\reviewer{The more likely possibility is that \method{} is able to correctly predict Li-enrichment when it is due to whatever mechanism preferentially enriches red clump stars.
If, as \citet{kumarDiscoveryUbiquitousLithium2020} posit, this mechanism is very common, it's especially likely that \method{} would capture its effects.}
Understanding which spectral features contain joint information with the Li doublet might indicate the mixing mechanism driving photospheric Li-enrichment. 
\reviewer{Notably, we identify few Li-enriched giants below the red clump.}

\begin{figure}
    \centering
    \includegraphics[width=\textwidth]{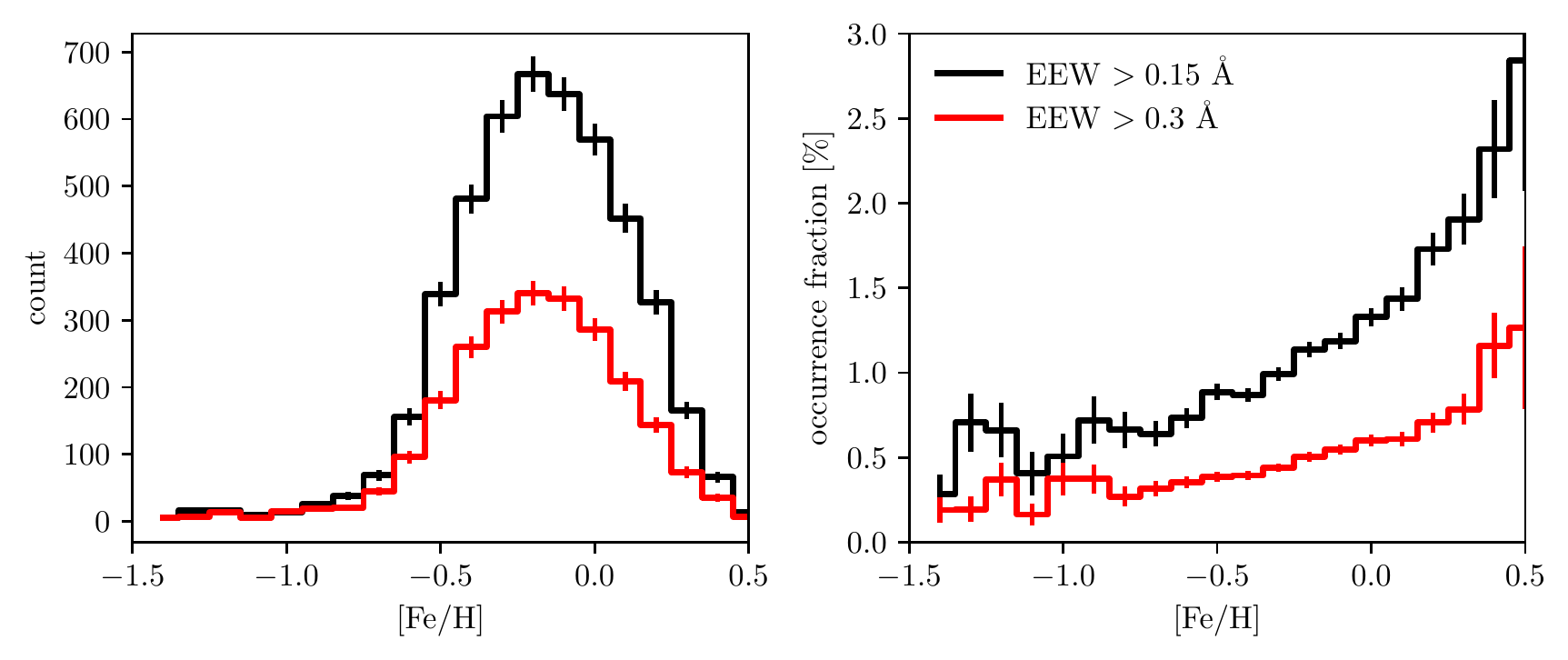}
    \caption{The raw count and occurrence fraction of Li-enriched giants as a function of metallicity. Errorbars are determined assuming Poisson uncertainty in raw counts. The fraction of Li-enriched giants increases smoothly with [Fe/H].}
    \label{fig:occurrencerate}
\end{figure}

In agreement with other recent work \citep{caseyTidalInteractionsBinary2019, martellGALAHSurveyLithiumrich2020, deepakStudyLithiumrichGiants2019}, we see that the occurrence rate of Li-enriched giants increases strongly with metallicity (Figure \ref{fig:occurrencerate}).
However, we see no evidence for a sharp increase in the occurrence fraction as [Fe/H] increases past the solar value, as noted by \citet{martellGALAHSurveyLithiumrich2020} and hinted at in the data of \citet{caseyTidalInteractionsBinary2019}.
Curiously, \citet{martellGALAHSurveyLithiumrich2020} found that a smooth increase on Li-enrichment was associated with RC stars, and that RGB stars only had enrichment ``turning on'' at super-solar metallicties.
\reviewer{We see a similar smooth increase in Li-enrichment with metallicity when plotting giants only above or only below the red clump.}
Those distinct trends are likely attributable to a multiplicity of enrichment mechanisms, further suggesting that our analysis is sensitive some mechanisms but not others.

\subsection{Young and pre-Main Sequence Stars}

Some Li-enriched stars are enriched by virtue of having not yet depleted their birth Li.
We are more likely to identify pre-main-sequence stars as Li-enriched than stars at any other evolutionary stage, with the enriched fraction exceeding 50\% for those with $\logg > 4$.
These stars are exhibit some of the largest EEWs (see Figure \ref{fig:kiel}), with some larger than 0.5 \AA{}, and roughly half have $H_\alpha$ emission, indicating magnetic activity.

\begin{figure}
    \centering
    \includegraphics{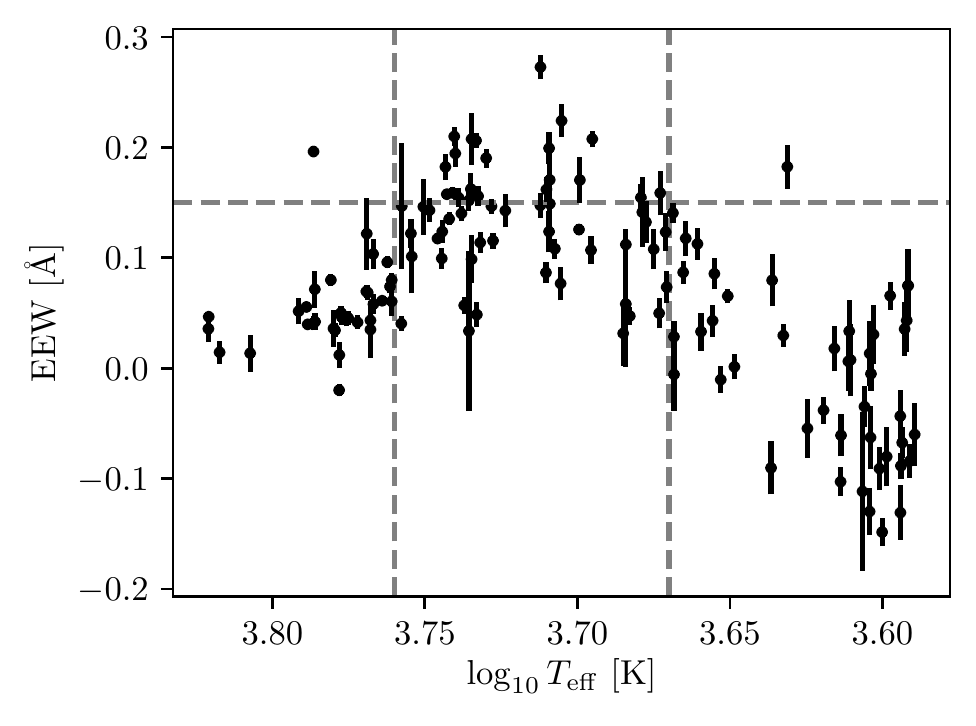}
    \caption{Li EEWs as a function of $\teff$ for the Pleiades stars observed by \lamost{}. 
    Note that the x-axis is reversed.
    The grey dashed line marks the threshold at which we consider a star Li-enriched.
    The decrease in EEW with increasing effective temperature is due to Li's low ionization potential.
    The decrease with decreasing effective temperature is because those stars have depleted more of their Li.}
    \label{fig:pleiades}
\end{figure}

We know from studies of open clusters that on and near the zero-age main sequence (ZAMS), the Li abundance is only mildly $\teff$-dependent for stars with $\teff \gtrsim 5000~\mathrm{K}$ ($\log_{10}\teff~[\mathrm{K}] \gtrsim 3.7$), but decreased quickly with decreasing $\teff$ below that value (e.g. \citealp{sestitoTimeScalesLi2005}).
Together with the fact that the 6708 \AA{} Li doublet is $\teff$-sensitive, this means that we primarily identify your main sequence stars of moderate effective temperature. 
These effects can be seen clearly in stars in the Pleiades (as identified with \gaia{} astrometry by \citealp{gaiacollaborationGaiaDataRelease2018}), the open cluster in which \lamost{} has observed the most stars.
Figure \ref{fig:pleiades} shows the EEW for each.
Dashed lines mark $\text{EEW} = 0.15~\text{\AA{}}$, our Li-enrichment threshold, and the temperatures within which Pleiades stars exceed this threshold: $4650 \lesssim \teff~[\mathrm{K}] \lesssim 5750$ ($3.67 \lesssim \log_{10} \teff~[\mathrm{K}] \lesssim 3.76$).
Reliably identifying young main sequence stars, even in a limited temperature range, can help to dissect the Galaxy's dynamical perturbations (e.g. \citealp{laporteBarResonancesLow2020}), identify and validate young exoplanetary systems (e.g. \citealp{mannTESSHuntYoung2020}), and potentially improve our understanding of star formation by identifying recently dispersed clusters (e.g. \citealp{fosterINSYNCIIVirial2015}).

\section{Discussion} \label{sec:discussion}
We have presented an unsupervised method, tangent space projection (TSP), for identifying stars which are high $X$-enriched, for any element $X$.
It uses the fact that most stellar spectra lie on a low-dimensional manifold, but those of chemically aberrant stars often don't.
We applied \method{} to the 6709 \AA{} Li doublet in \lamost{} DR5, identifying \nstars{} Li-enriched stars.


\method{} is applicable to any homogeneous catalog of spectra.
Its most useful application is to blended features and those for which physical modelling remains a challenge.
For low-resolution surveys like \lamost{}, essentially all spectral features \reviewer{are blended to some extent}, but medium- and high- resolution surveys, such as 
\gaia{} DR4, \textsl{RAVE} \citep{steinmetzSixthDataRelease2020, caseyRAVEonCatalogStellar2017},  \apogee{} \citep{majewskiApachePointObservatory2017, garciaperezASPCAPAPOGEEStellar2016}, and Sloan V \citep{kollmeierSDSSVPioneeringPanoptic2017} also contain many such features, e.g. the recently noted Ce and Nd lines \citep{hasselquistIDENTIFICATIONNEODYMIUMAPOGEE2016, cunhaAddingSProcessElement2017} in \apogee{}.
In the disk, we expect to find few stars with unusual enrichment patterns, but they have to potential to give us unique insight into nucleosynthetic processes \citep{weinbergChemicalCartographyAPOGEE2019a}.
In the halo, the chemical signatures are strongly linked to the dynamical history of the Milky Way (e.g. \citealp{jiSouthernStellarStream2020, dasAgesKinematicsChemically2020, naiduEvidenceH3Survey2020}).

\reviewer{
\subsection{Blended lines} \label{sec:ba}
\begin{figure}
    \centering
    \includegraphics[width=\textwidth]{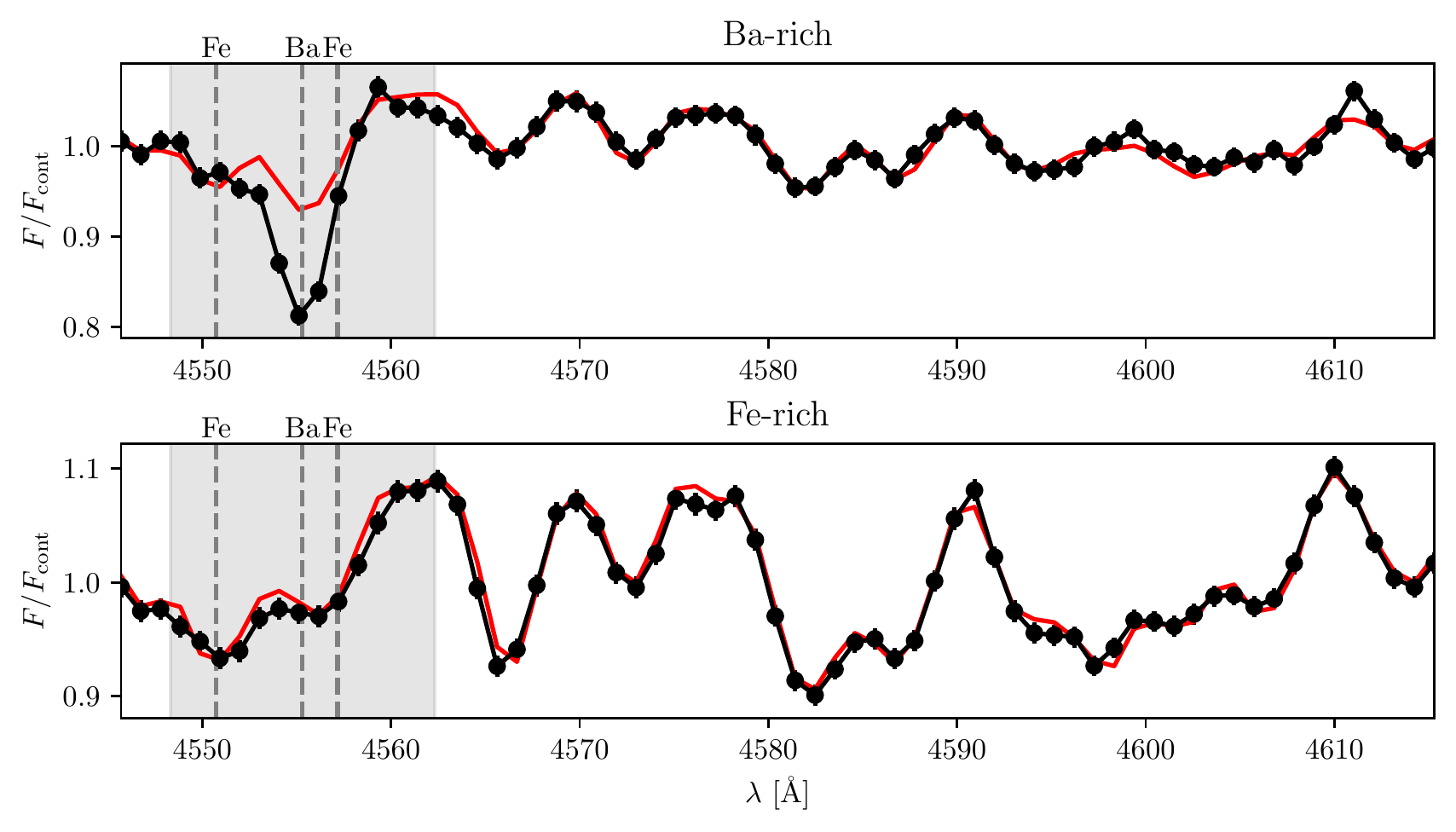}
    \caption{A portion of a spectrum of an Ba-rich star (\textbf{top}; \lamost{} J120744.31+172831.6) and an Fe-rich star (\textbf{bottom}; \lamost{} J093835.58-064404.6) as observed by \lamost{}.  \method{} correctly imputes the masked region (marked in grey) for the Fe-rich star, but not the Ba-rich star.  Note that the continuum is underestimated to different degrees in each, so the y-axes are not directly comparable.}
    \label{fig:baspectra}
\end{figure}
The $6708~$\AA{} Li doublet is blended with a weak Fe I line, which does not pose an analysis problem except in very metal-rich stars.
To demonstrate that TSP can still effectively identify chemical outliers in the presence of stronger blended features, we masked and imputed the region surrounding the 4555 \AA{} Ba line (with all other Ba lines masked), which is blended with two Fe lines. 
Figure \ref{fig:baspectra} shows a portion of the spectrum for a Ba-enriched (and Ba-rich) star, and for a Fe-rich, Ba-normal star.
As expected, \method{} correctly imputes the masked region when there is strong absorption due to the Fe line, but not when there is strong absorption from the Ba line.
Estimates of abundances from supervised models are often unreliable for extreme abundance values.
A complete analysis of blended features should also take blends into account when examining the imputed residuals.
For example, were the 6708 \AA{} Li doublet blended with a stronger line, we would have included absorption due to this feature as a contaminant model (section \ref{sec:flagging}).
}

\subsection{Algorithm}
We see the strengths of \method{} not in sophistication, but in simplicity and suitability to the problem addressed.
Further simplifications would come at the expense of predictive accuracy, as discussed in section \ref{sec:hyperparams}.

There are several elaborations on \method{} potentially appropriate for future work.
We found that predictive accuracy increase with the number of stars in the reference set, but we truncated our parameter search at a reference set of 30,000 stars for the sake of speed.
Identifying the nearest spectral neighbors, currently the most computationally expensive step, could be accelerated in a variety of ways, e.g. by making a pre-pass with the spectra in compressed form.
Presumably, the presence of Li-enriched stars in our reference set prevented us from identifying a fraction of enriched stars.
Removing Li-enriched reference stars from the reference set via iterative application  of \method{} could address this potential problem and give more complete results.

Equation (\ref{eq:proj}) is a form of unregularized regression.
Ideally, all projection weights will be small, since large weights correspond to a part of the tangent space that doesn't overlap with the spectral manifold.
We have found that this is true for our data, but a regularized form of Equation (\ref{eq:proj}) could help ensure that is is more often the case.

In this work, we separate calculation of the imputed prediction, $\m\f^\text{pred}$, from examining residuals to identify strong absorption.
Jointly fitting for EEW simultaneously with projection onto the approximate tangent space would be more  principled, and potentially more effective.
Using a probabilistic or robust form of PCA (e.g. \citealp{bishopPatternRecognitionMachine2006} chapter 12) could give a more principled estimate of the tangent space.
Other neighbor selection schemes are also possible, such as using all neighbors within a given distance hypersphere. An adaptive selection scheme could potentially adjust dynamically for the changing density of the reference set across the manifold.

In section \ref{sec:hyperparams}, we showed that the imputed prediction in the region of the Li doublet is remarkably insensitive to the choice of $k$ and $q$.
In fact, we found this to be the case even when we didn't mask the Li doublet region, presumably because the number of pixels outside the doublet region dwarf the number of Li-sensitive pixels, and dominate equation \ref{eq:proj}.
In addition, if the subspace dimension, $q$, is small, the model will have too few degrees of freedom to capture Li variation independently from other abundances.

One might ask what advantages \method{} has over any established nonlinear dimensionality reduction algorithm to lossily compress and decompress spectra.
While such approaches would likely work, our censor-and-predict scheme is not naturally supported by these tools.
Even if censoring turns out to be unnecessary for real data, we avoid a large computational optimization problem because we don't need a global and continuous low-dimensional basis.
We incorporate uncertainty during the prediction step, and the simplicity of \method{} means that a justified uncertainty estimate is easy to compute.

\subsection{What could possibly go wrong?}
Here we list situations and ways in which \method{} can fail to identify $X$-enriched or $X$-depleted stars.
Visualizing all steps of the analysis is the best way to diagnose these kinds of problems.
\begin{itemize}
   \item \textsl{The manifold is not well-sampled.} If there are not enough points in the reference set, the $k$-nearest points may ``jump across a wrinkle in the manifold''.
   This will result in a poor fit to both the masked and unmasked pixels.
   \item \textsl{The manifold is sampled very non-uniformly.} In general, the reference set will not be uniformly distributed across the manifold, meaning that the above issue can arise in some regimes, but not others.  Using an adaptive method to pick the neighborhood size, $k$, may help in these situations.
   \item \textsl{The mask is wrong, i.e. the abundance effects the spectrum in many places}.  If the abundance $X$ affects stellar spectra in regions outside of the mask, \method{} may be able to predict the masked pixels even of chemically aberrant stars.
   This may result in identified outliers disappearing for large subspace dimension, $q$, when the model has enough degrees of freedom to capture $A(X)$ as an independent factor of variation.
   \item \textsl{The line model is wrong.} If the line location or profile are wrong, some enriched stars may fail to be identified, and EEWs may be misestimated.  This will typically be identifiable by visualizing residuals, $\m\rr$, and best-fit line models, $\alpha^*\mv$.
   \item \textsl{The contaminant models are incomplete or too eager.} Similarly, an unaccounted-for or misspecified contaminant can cause stars to be misclassified. Again, visualization is the best way to identify this situation.
   \item \textsl{The method underperforms relative to a supervised model}. Finally, a supervised data-driven model, or a completely \emph{ab-initio} physical model will be a more appropriate choice, in cases where the physical models are fast and accurate or training labels are abundant, precise, and accurate.
\end{itemize}
We believe that this work is not hampered significantly by any of these issues.  
Visualization indicates that the model fits the data closely and can accurately predict held-out data.
The tests described in section \ref{sec:hyperparams} shows that the typical predictive accuracy is better than the nominal measurement uncertainty.
The Li doublet's profile is dominated by instrumental dispersion, and its profile is well-described by a Gaussian, and its wavelength precisely known.

\section{Conclusions}
Our chief findings are the following:
\begin{itemize}
    \item We introduce \method{}, a method for imputing data using ideas from manifold learning which can be used to identify stars enriched in a given element from their spectra without a physical model.
    \item We apply TSP to the 6708 \AA{} Li doublet in \lamost{}, identifying \nstars{} Li-enriched stars.
    \item We examine the abundances of Li-enriched stars near the 1.5 $M_\odot$ subgiant branch, the regime where Li-enrichment is thought to be most likely to be due to planet accretion. We find that Li-enriched and Li-normal subgiants have nearly identical individual abundance distributions in the 10 elements examined and are thereby not distinguished by any signature of potential engulfment in other abundances.
    \item We do not see a sharp increase in the fraction of Li-enriched red giants at solar [Fe/H] and we identify few Li-enriched red giants with log g above that of the red clump.
    Furthermore, we see no surplus of Li-rich giants on the red clump, in contraction to prior studies.
    This suggests that TSP are sensitive to a different set of enrichment mechanisms than abundance-based searches. 
    \item Using observations of stars in the Pleiades, we demonstrate that we reliably identify young main sequence stars with $4650 \lesssim \teff~[\mathrm{K}] \lesssim 5750$ as Li-enriched.
\end{itemize}

\software{matplotlib \citep{caswellMatplotlibMatplotlibREL2020}}

\section*{Acknowledgments}
The authors thank the Milky Way Stars group at Columbia.
AJW is supported by the National Science Foundation Graduate Research Fellowship under Grant No. 1644869.
MKN is in part supported by a Sloan Research Fellowship.

\bibliography{xref}

\end{document}